\documentclass[aps,pra,amsfonts,amsmath,amssymb,floatfix,reprint,superscriptaddress]{revtex4-2}
\usepackage{graphicx, hyperref}

\newcommand*{\avg}[1]{\langle #1 \rangle} % for average
\newcommand*{\abs}[1]{\left| #1 \right|} % for absolute value
\newcommand{\ket}[1]{\left| #1 \right>} % for Dirac bras
\newcommand{\bra}[1]{\left< #1 \right|} % for Dirac kets
\newcommand{\braket}[2]{\left< #1 \vphantom{#2} | #2 \vphantom{#1} \right>} % for Dirac brackets
\DeclareMathOperator{\Tr}{Tr}
\DeclareMathOperator{\var}{var}

\begin{document}

\title{Thermalisation in a Bose-Hubbard dimer with modulated tunneling}

\author{R. A. Kidd}
\email{ryan.kidd@uq.edu.au}
\affiliation{School of Mathematics and Physics, University of Queensland, Brisbane, Queensland 4072, Australia.}
\author{A. Safavi-Naini}
\affiliation{ARC Centre of Excellence for Engineered Quantum Systems,
School of Mathematics and Physics, University of Queensland, Brisbane, QLD 4072, Australia}
\affiliation{School of Mathematics and Physics, University of Queensland, Brisbane, Queensland 4072, Australia.}
\author{J. F. Corney}
\affiliation{School of Mathematics and Physics, University of Queensland, Brisbane, Queensland 4072, Australia.}

%-----------------------------------------------------------------------
\date{\today}
%------------------------------------------------------------------------

\begin{abstract}

The periodically modulated Bose-Hubbard dimer model offers an experimentally realizable and highly tunable platform for observing the scrambling of quantum information and the apparent thermalisation of isolated, interacting quantum many-body systems. In this work we apply the fidelity out-of-time-order correlators in Ref.~\cite{Lewis-Swan2019} to establish connections between thermalisation in Floquet system, the exponential growth of FOTOCs as quantified by a non-zero quantum Lyapunov exponent, and the underlying classical transition from regular to chaotic dynamics in the dimer. Moreover, we demonstrate that a non-zero quantum Lyapunov exponent can also be inferred from measures quantifying the delocalisation of the Floquet modes of the system such as the Shannon entropy, which approaches unity if the system thermalises to the periodic Gibbs ensemble prediction.

\end{abstract}


%*******************************************

\maketitle

\section{Introduction}

Periodically driven quantum systems can be engineered to display out-of-equilibrium exotic many-body phenomena, such as dynamical localisation and transition from integrable to chaotic dynamics as a result of varying the driving parameters~\cite{DAlessio2014}. Hence these systems can facilitate one of the current research fronts in quantum many-body dynamics: thermalisation and its connection to the dynamics of quantum information, correlations, and quantum entanglement.

The connection between thermalisation and the dynamics of quantum information is manifest in \textit{quantum scrambling}, in which two initially commuting operators become rapidly delocalised and non-commutative, as quantified by out-of-time-order correlators (OTOCs)~\cite{Hayden2007, Sekino2008, Shenker2014, Hosur2016, Swingle2016, Maldacena2016}. However, OTOCs are not easily accessible in most quantum simulation platforms.

Time-reversal protocols similar to Loschmidt echoes have been used to measure OTOCs in an Ising spin system~\cite{Garttner2017}, a nuclear magnetic resonance quantum simulator~\cite{Li2017} and a nuclear spin system in a natural crystal~\cite{Wei2019}. For some systems, time reversal protocols can be difficult to implement and alternative OTOC measurement schemes have been developed, utilising interferometric protocols~\cite{Yao2016}, auxiliary degrees of freedom~\cite{Landsman2019}, statistical correlations~\cite{Vermersch2019, Joshi2020}, operator variance measurements~\cite{Lewis-Swan2019, Pilatowsky-Cameo2020} and operator eigenbasis measurements~\cite{Chavez-Carlos2019, Blocher2020}.

Here we consider OTOC protocols for periodically driven systems, which can be described with a Floquet formalism~\cite{Eckardt2015}. An isolated, nonintegrable Floquet system thermalises to the equivalent of an infinite-temperature state, in the sense that its few-body observables approach the values predicted by the maximal-entropy `diagonal ensemble'~\cite{DAlessio2014, DAlessio2016, Seetharam2018}. Such thermalisation in isolated quantum systems is predicted by the eigenstate thermalisation hypothesis (ETH), which pertains to the statistical behaviour of the eigenvalues, or in the case of a Floquet system, the quasienergies~\cite{DAlessio2016}.

In this paper we demonstrate the use of the \textit{fidelity} out-of-time-order correlator (FOTOC)~\cite{Lewis-Swan2019} for studying the manifestation of chaos in a system of ultra-cold bosons in a double-well potential with periodically modulated tunnelling~\cite{Kidd2019}. Our proposal can be immediately realised experimentally and highlights the utility of FOTOC dynamics for characterizing scrambling and thermalisation in Floquet systems. This paper is organized as follows: we introduce the physical system of periodically modulated ultracold lattice-bound bosons followed by a brief introduction of Floquet theory. We then introduce FOTOCs in the context of quantum scrambling, showing distinct behaviours that can be linked with the semiclassical regular and chaotic regimes. Finally we compare the behaviour of FOTOCs to statistical indicators of thermalisation, such as level-spacing parameters and spectral delocalisation of Floquet modes.

\section{The Bose-Hubbard dimer}

We consider a system of ultra-cold bosons in a double-well potential within the tight-binding approximation. At sufficiently low temperatures, the system is described by a two-mode model, which upon introducing the ladder operators $\hat{a}_j$, $\hat{a}_j^\dagger$ associated with the occupation in the $j$th well, takes the form of a two-site Bose Hubbard model.

The Hamiltonian governing the system dynamics can be written
\begin{equation} \label{eq:H}
    \hat{H} = 2U \hat{S}_z^2 - 2J\hat{S}_x,
  \end{equation}
where we have introduced the pseudo-angular-momentum operators,
\begin{eqnarray}
    \hat{S}_x &=& \frac{\hat{a}_2^\dagger \hat{a}_1 + \hat{a}_1^\dagger \hat{a}_2}{2}, \nonumber \\
    \hat{S}_y &=& \frac{\hat{a}_2^\dagger \hat{a}_1 - \hat{a}_1^\dagger \hat{a}_2}{2i}, \nonumber \\
    \hat{S}_z &=& \frac{\hat{a}_2^\dagger \hat{a}_2 - \hat{a}_1^\dagger \hat{a}_1}{2},
\end{eqnarray}
with $\hat S_i$ satisfying the commutation relations $[\hat{S}_\alpha, \hat{S}_\beta] = i \epsilon_{\alpha \beta \gamma} \hat{S}_\gamma$.
Here $J$ is the tunneling strength and $U$ is the on-site interaction energy.

In the noninteracting limit, this quantum dimer exhibits Rabi oscillations akin to a pseudospin-1/2 particle. With increasing interaction strength $U/J$, the unmodulated dimer undergoes a pitchfork bifurcation at the critical interaction strength $U_c/J=1/N$~\cite{Milburn1997}. This bifurcation corresponds to the onset of self-trapping, which has been experimentally observed~\cite{Albiez2005}.

In addition to the above phase transition, periodic modulation of the coupling rate $J(t)= J_0 + \mu \cos{(\omega t)}$ introduces chaotic behaviour to the dimer model~\cite{Milburn1997chaos}.
The chaotic behaviour is manifest in the semiclassical dynamics of the system and can be studied at the mean-field level using the expectation values $(x,y,z) \equiv (\avg{\hat{S}_x},\avg{\hat{S}_y},\avg{\hat{S}_z})$. Under particle conservation, the resulting phase space is spanned by two parameters $\vec r = (z,\phi)$, where $z=\avg{\hat{S}_z}$ and $\phi=-\arg{\left( \avg{\hat{S}_x}+i\avg{\hat{S}_y} \right)}$.

\begin{figure}[t]
  \includegraphics[width=\columnwidth]{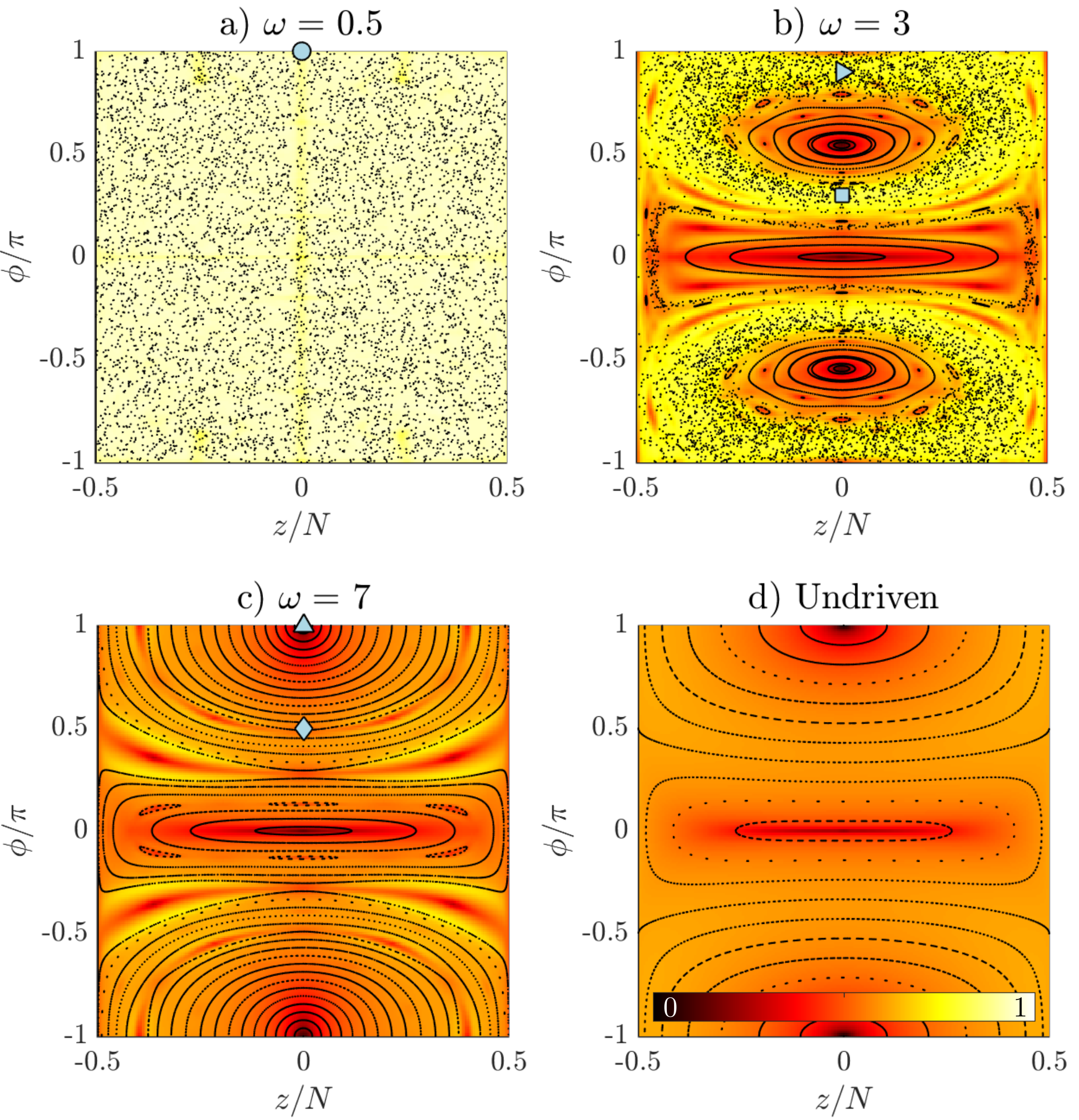}
  \caption{\label{fig:poincare_S}Poincar\'e sections of the semiclassical dynamics (black points), overlaid on colour maps of coherent-state Shannon entropy in the basis of Floquet modes, for a range of driving parameters, as indicated, from (a) fully chaotic to (d) fully regular. Blue markers indicate initial conditions for Fig.~\ref{fig:FOTOC}. Parameters are $NU=-1$ and $J(t)=1+1.5\cos{(\omega t)}$. The colour map corresponds to the Shannon entropy of the $N=1000$ coherent state centred at that point in phase space, scaled by $\mathcal{S}_\text{max} = \log{(N+1)}$ to give the range $[0,1]$. Chaotic regions are associated with large Shannon entropy, whereas in stable regions the Shannon entropy tends to be small, particularly near the stable centres}
\end{figure}

In Fig.~\ref{fig:poincare_S} we use stroboscopic Poincar\'e sections, with trajectories plotted at intervals of the modulation period, to illustrate the regular-to-chaotic transition as a function of the modulation frequency. We tune between fully regular (no driving, or effectively $\omega \rightarrow \infty$) to fully chaotic ($\omega = 0.5$), passing through a region of mixed phase space with regions of chaos and regions of regular dynamics.

Chaotic behaviour in classical systems is diagnosed by the Lyapunov exponent, which characterises the rate of exponential growth of initially close trajectories:
\begin{equation}
    \lambda = \lim_{t\to\infty} \lim_{\delta \vec{r}_0 \to 0}  \frac{1}{t} \ln{\abs{\abs{ \frac{\delta \vec{r}(t)}{\delta \vec{r}_0} }}},
\end{equation}
Thus, the regular and chaotic regions in the semiclassical phase space correspond to zero and non-zero values of $\lambda$, respectively. The chaotic dynamics of the modulated dimer was explored in Ref.~\cite{Kidd2019}, where the Lyapunov exponent was correlated with various probes of chaotic behaviour in the corresponding quantum dynamics.

In the quantum regime, the dimer model is a numerically tractable model which allows us to explore connections between classical chaos in periodically modulated systems, the dynamics of quantum information as characterized by the growth of OTOCs, and thermalisation. In order to establish these connections we utilise Floquet theory to study the dynamics generated by Eq.~\eqref{eq:H} under periodic modulation of $J(t)$. This periodic driving scheme can be readily implemented in bosonic lattice systems by modulating optical lattice depth~\cite{Bloch2005, Eckardt2017}, and interparticle interaction strength can be controlled via Fesbach resonance~\cite{Chin2010}. In these systems, the configurable trapping potential and interparticle interactions allows for arbitrary time-periodic Floquet driving schemes to be applied to ultracold atom systems~\cite{Eckardt2017}.

The functional form of the modulation creates a time-periodic Hamiltonian $\hat{H}(t) = \hat{H}(t + T)$ with driving period $T=2\pi/\omega$. Floquet systems are characterised by the complex eigenvalues of the time-evolution operator over one period, which is given by
\begin{equation}
    \hat{U}(T,0) \ket{\phi_\alpha} = e^{-i\epsilon_\alpha T} \ket{\phi_\alpha},
\end{equation}
where we have set $\hbar =1$. Here $\ket{\phi_\alpha}$ is a Floquet mode and $\epsilon_\alpha$ is the associated quasienergy.

In general, the Floquet Hamiltonian, $\hat{H}_F$, is defined by
\begin{equation} \label{eq:U_F}
    \hat{U}(T,0) = e^{-i \hat{H}_F T}.
\end{equation}
For sufficiently large driving frequency (larger than the bandwidth of the time-dependent Hamiltonian~\cite{DAlessio2014}), we can find a perturbative approximation to $\hat{H}_F$ using the Floquet-Magnus expansion~\cite{Bukov2015},
\begin{equation} \label{eq:H_highf}
    \hat{H}_\text{eff} = \sum_{j=0}^\infty T^j \hat{\Omega}_{j}=\hat \Omega_0 + \hat \Omega_2 T^2 + O(T^3),
 \end{equation}
with
 \begin{eqnarray}
    \hat{\Omega}_0 &=& 2U \hat{S}_z^2 - 2J_0 \hat{S}_x, \\
    \hat{\Omega}_2 &=& \frac{2\mu U^2}{\pi^2} \left( \hat{S}_x + 4\hat{S}_z \hat{S}_x \hat{S}_z \right)
  + \frac{\mu U}{\pi^2} \left(\mu - 4J_0 \right) (\hat{S}_y^2 - \hat{S}_z^2)\nonumber.
\end{eqnarray}
For large driving frequencies the contribution from the $\hat{\Omega}_2$ term is negligible, in which case $\hat{H}_\text{eff}$ reduces to the time-averaged, or unmodulated, Hamiltonian~\eqref{eq:H}. We note that the effective Hamiltonian is valid only when the quasienergies $\epsilon_\alpha$ do not exhibit level-repulsion consistent with random matrix theory~\cite{DAlessio2014}. In other words, the breakdown of the Magnus expansion is an indication of the chaotic behaviour in the quantum dynamics. In this paper, we generally determine the Floquet quasienergies and modes numerically, using QuTiP~\cite{Johansson2012, *Johansson2013}.

Further insight into the breakdown of the Magnus expansion, and hence the failure of the effective Hamiltonian approximation~\cite{Santos2010_1}, can be obtained analysis of the delocalisation of the Floquet modes in the basis of effective Hamiltonian eigenstates~\cite{DAlessio2014}. The Shannon entropy, $\mathcal{S}$, quantifies this delocalisation~\cite{Santos2010_1} and is defined as
\begin{equation}
    \mathcal{S}_n = \sum_m \abs{c_n^m}^2 \ln{\left(\abs{c_n^m}^2 \right)},
\end{equation}
where $\abs{c_n^m}^2 = \braket{\psi_n}{\phi_m}$ is the overlap between the Floquet mode $\ket{\phi_m}$ and the effective Hamiltonian eigenstate $\ket{\psi_n}$, satisfying $\hat{H}_\text{eff} \ket{\psi_n} = E_n \ket{\psi_n}$.

A related delocalisation measure can be obtained by calculating the Shannon entropy of \textit{coherent states} in the basis of Floquet modes, which yields a distribution over phase space. Fig.~\ref{fig:poincare_S} compares the delocalisation of coherent states across phase space in the basis of Floquet modes to the corresponding semiclassical phase space features. Chaotic phase space regions are accompanied by high delocalisation, while regular regions are accompanied by low delocalisation. Coherent states centred on phase space fixed points exhibit near-minimal delocalisation.

States that are sufficiently delocalised in the Floquet modes, as indicated by Shannon entropy close to the circular orthogonal ensemble prediction, $\mathcal{S}_\text{COE} \approx \ln{\left[0.48(N+1)\right]}$~\cite{DAlessio2014}, are expected to synchronise to the modulation by asymptotically approaching a periodic Gibbs ensemble, a limit cycle with a period given by the modulation period~\cite{Lazarides2014_1}.

\section{Scrambling}

The quantum Lyapunov exponent $\lambda_Q$ characterises the rate at which quantum information is scrambled, that is, where initially local quantum information spreads over the degrees of freedom of the system. Previous studies have established that the presence of classical chaos, characterized by $\lambda>0$, indicates that in the corresponding quantum system $\lambda_Q>0$~\cite{Lewis-Swan2019, Blocher2020, Xu2020}. In a quantum system, $\lambda_Q$ can be extracted from the measurement of an OTOC,
\begin{equation}
  C(t) = \avg{|[\hat{W}(t),\hat{V}]|^2}, \qquad
  \hat{W}(t) = \hat{U}^\dagger(t) \hat{W} \hat{U}(t),
\end{equation}
where $\hat{W}$ and $\hat{V}$ are two operators and $C(t)$ characterizes the non-commutativity of $\hat{V}$ and and $\hat{W}$ at a later time $t$. For quantum many-body systems with a chaotic classical limit, the OTOC grows exponentially as $C(t) \sim e^{\lambda_Q t}$~\cite{Hayden2007, Sekino2008, Shenker2014, Maldacena2016, Shen2017, Bentsen2019_PNAS}.

In the following we focus on the behaviour of a specific family of OTOC, the fidelity OTOC (FOTOC), which has been used to probe the connections between chaos, scrambling, and thermalisation in the Dicke model~\cite{Lewis-Swan2019}, and has been experimentally implemented in a quantum simulator of the all-to-all Ising model composed of a crystal of hundreds of ions~\cite{Garttner2017}. To generate the FOTOC, we choose $\hat{W} = e^{i \delta \hat{w}}$, where $\hat{w}$ is the generator of an arbitrary rotation, and $\hat{V} = \ket{\psi_0}\bra{\psi_0}$, with $\ket{\psi_0}$ the initial state of the system.

For small perturbations, $\delta \ll 1$, the FOTOC reduces to the variance of $\hat{w}$ since $C(t)\approx \delta^2 \var{[\hat{w}(t)]} + O(\delta^3)$~\cite{Schmitt2019}. This allows for a simple experimental implementation of the FOTOC since the variance of the particle-number difference, $\hat{S}_z$, is directly measurable experimentally, even to single-particle precision~\cite{Stroescu2015}. Indeed, the variance along any axis can be experimentally measured, through appropriate rotations on the Bloch sphere around $\hat{S}_x$ (interwell-tunnelling)~\cite{Tomkovic2017} and $\hat{S}_y$ (relative energy difference). For the rest of this work we restrict ourselves to this regime.

To provide a local probe in a mixed phase space featuring both regular and chaotic dynamics, we calculate FOTOCs for arbitrary initial coherent states on the Bloch sphere. Furthermore, to ensure the FOTOC is initially zero, we choose the operator $\hat{w}$ such that the initial coherent state is an eigenstate~\cite{Lewis-Swan2019, Blocher2020}. Thus for a spin coherent state centred on azimuthal angle, $\phi$, and polar angle, $\theta$, we choose
\begin{equation} \label{eq:local_op}
    \hat{w} = \cos(\phi)\sin(\theta) \hat{S}_x + \sin(\phi)\sin(\theta) \hat{S}_y + \cos(\theta) \hat{S}_z.
\end{equation}

\begin{figure*}[t]
    \includegraphics[width=\textwidth]{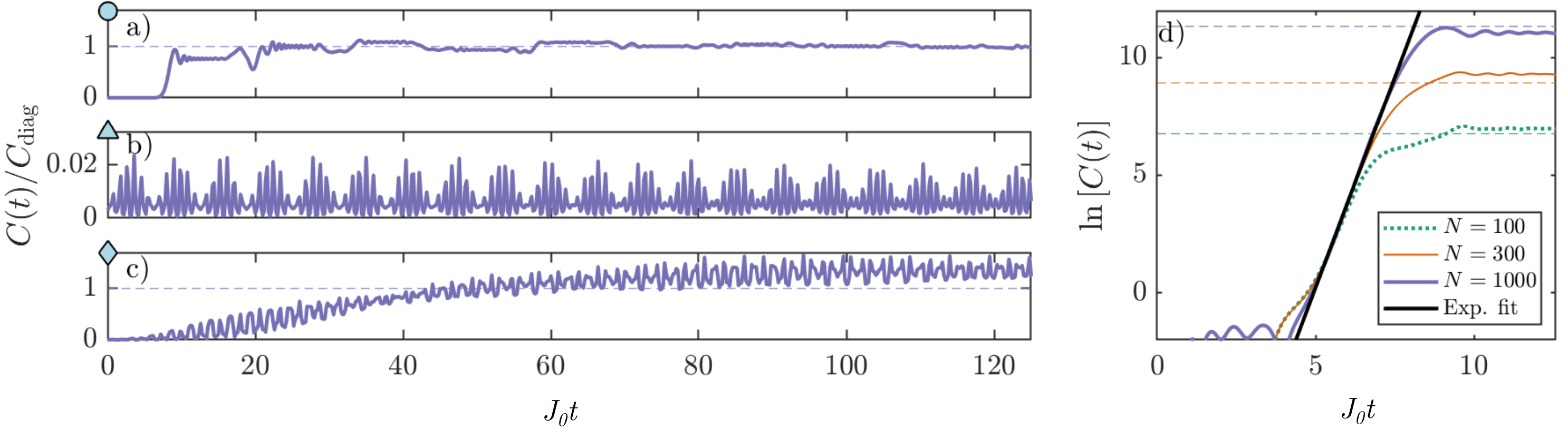}
    \caption{\label{fig:FOTOC}Fidelity out-of-time-order correlator (FOTOC) for $N=1000$ initial coherent state in: a) chaotic phase space ($\omega = 0.5$); regular phase space ($\omega = 7$) on a stable fixed point; and c) regular phase space ($\omega = 7$) away from a stable fixed point. Blue markers correspond to FOTOC initial conditions shown in Fig.~\ref{fig:poincare_S}. d) FOTOCs for various particle numbers with parameters as in (a) and $\delta = 10^{-2}$.  Here the dotted green, thin orange and thick purple lines represent $N=100,300,1000$, respectively; exponential fit is $\lambda_Q = 3.58(3)$ in units of $J_0^{-1}$. Horizontal dashed lines are diagonal ensemble predictions. Parameters for all plots are $NU = -1$, $J(t) = J_0 + 1.5\cos{(\omega t)}$.}
\end{figure*}

In Fig.~\ref{fig:FOTOC} we plot the FOTOC dynamics for parameters in the chaotic (panels a and d) and regular (panels b and c) phase spaces for $N=1000$. We find that in the regime where the system is classically chaotic, after an initial delay time, the FOTOCs grow exponentially before plateauing at a finite value predicted by the diagonal ensemble for an infinite-temperature state, in accordance with the Floquet eigenstate thermalisation hypothesis prediction. As expected~\cite{Lewis-Swan2019}, both the saturation time (Ehrenfest time) and the saturation value grow with increasing $N$. In Fig.~\ref{fig:FOTOC}b, the FOTOC for a stable fixed point in phase space remains small-valued and appears quasiperiodic, but does not synchronise with the modulation period. In Fig.~\ref{fig:FOTOC}c, the FOTOC in regular phase space away from a fixed point grows slowly and reaches a large-valued quasiperiodic limit after about 10 Ehrenfest times. Unlike the chaotic FOTOC, the regular FOTOC limit has large-amplitude oscillations and does not approach the infinite-temperature diagonal-ensemble prediction. The average value of the quasiperiodic limit is greater than the infinite-temperature diagonal-ensemble prediction due to the state's ring-like $Q$-distribution.

\begin{figure}[ht]
    \centering
    \includegraphics[width=0.6\columnwidth]{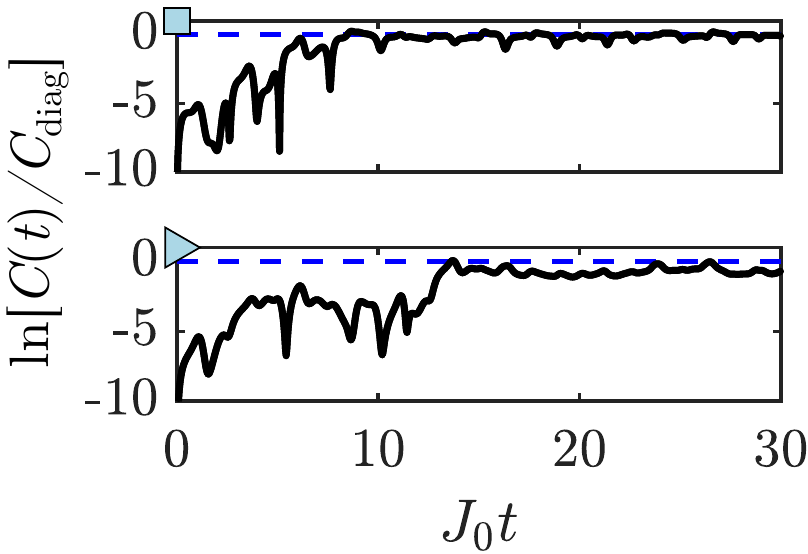}
    \caption{\label{fig:FOTOC2}Dynamics of FOTOC logarithms scaled by diagonal ensemble prediction, $C_\text{diag}$, for reference states indicated by blue markers in Fig.~\ref{fig:poincare_S}. Parameters are $N = 1000$, $NU = -1$ and $J(t) = J_0 + 1.5\cos{(5t)}$.}
\end{figure}

It is worth noting that the behaviour of the FOTOC can be a mixture of the behaviours described above if the state has overlap with both regular and chaotic regions of the phase space. We indicate two such states using blue markers in Fig.~\ref{fig:poincare_S}b and plot their respective FOTOCs in Fig.~\ref{fig:FOTOC2}. Hence, to observe pure exponential growth of a FOTOC it is important to tune the Hamiltonian parameters such that the semiclassical phase diagram has well-defined chaotic and regular regimes that are wide enough with respect to the quantum noise of the initial state.

\section{Thermalisation}

The exponential growth of the FOTOC discussed above closely resembles the behaviour of the FOTOC in the Dicke model~\cite{Lewis-Swan2019}. Thus we expect the previously established relationship between thermalisation and chaos to hold. In the case of the modulated Bose-Hubbard dimer, however, the eigenstate thermalisation hypothesis (ETH) predicts that the system will approach an infinite-temperature state~\cite{DAlessio2014}, of the form $\hat{\rho} = \mathbb{I}/(N+1)$. The variance of an arbitrary local operator~\eqref{eq:local_op} in this limit is
\begin{eqnarray}
    \var{(\hat{w})} &=& \Tr{(\hat{\rho} \hat{w}^2)} - \Tr{(\hat{\rho} \hat{w})}^2 \nonumber \\
    &=& \Tr{(\hat{w}^2)}/(N+1) - \Tr{(\hat{w})}^2/(N+1)^2 \nonumber \\
    &=& N(N+2)/12,
\end{eqnarray}
where we have used $\Tr{(\hat{S}_\alpha)} = 0$ and $\Tr{(\hat{S}_\alpha \hat{S}_\beta)} = \delta_{\alpha\beta} N(N+1)(N+2)/12$. Therefore, $C_\text{diag} = \delta^2 N(N+2)/12$.

The approach to the infinite-temperature thermodynamic limit is evident in Fig.~\ref{fig:FOTOC}d, where the average values of the FOTOCs at long times approach the diagonal ensemble predictions shown with dashed lines.

The equilibration to infinite temperature is a result of the properties shared between the Floquet operator~\eqref{eq:U_F} and random matrices~\cite{DAlessio2014}. This association with random matrices is revealed in the quasienergy statistic, $\avg{r}$. Level-spacing statistics more generally have long been used as the defining characteristic of chaos in quantum systems~\cite{Haake2010}. In Floquet systems, the spacings between adjacent quasienergies within the same symmetry class of the Hamiltonian play the role of the eigenvalue spacing for time-independent systems. Hence we expect circular ensemble statistics when our system equilibrates to infinite temperature, and Poissonian statistics otherwise~\cite{DAlessio2014}.

\begin{figure}[t]
    \centering
    \includegraphics[width=\columnwidth]{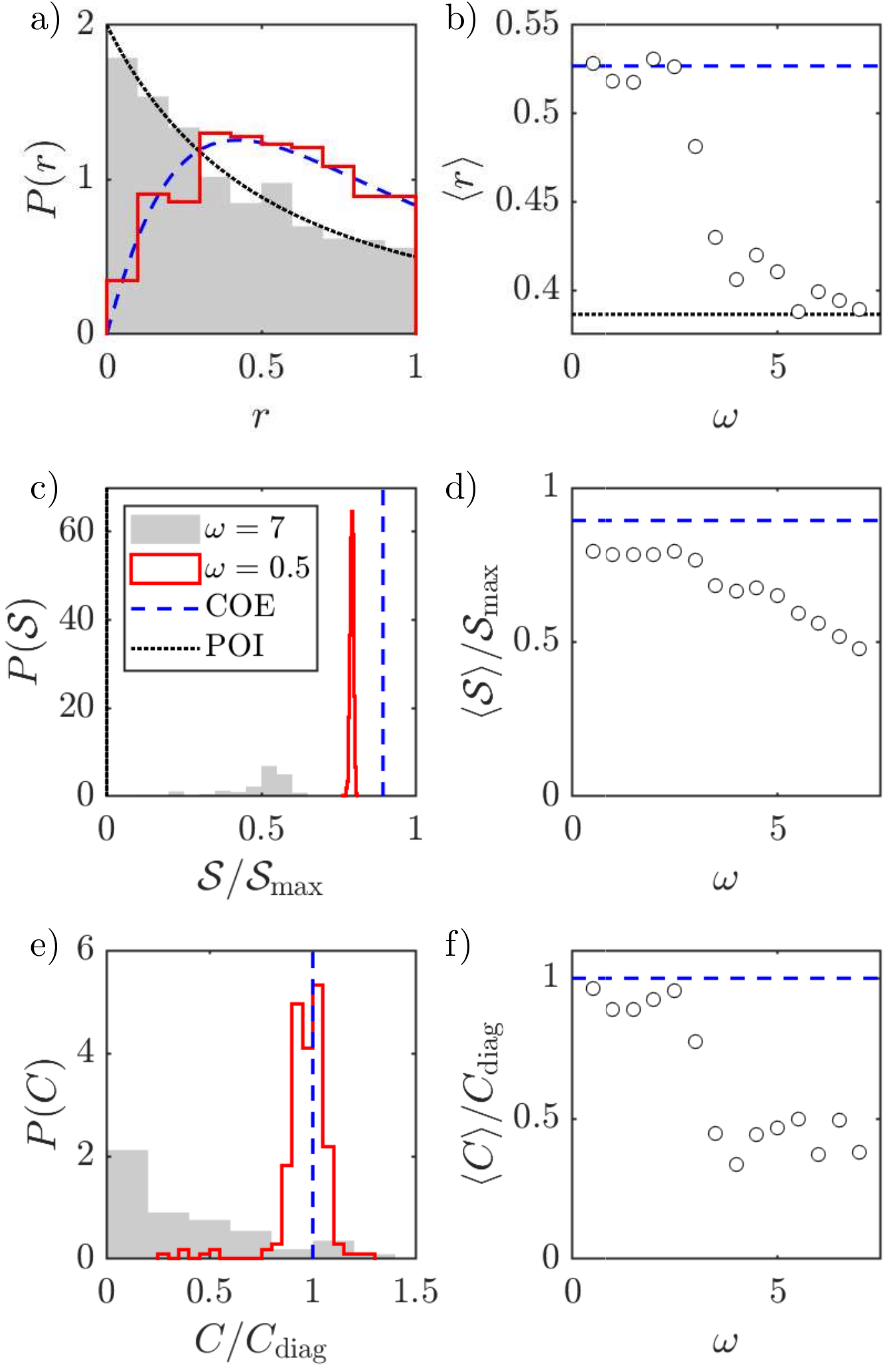}
    \caption{\label{fig:spectrum}a) Quasienergy level spacing distributions for widely integrable ($\omega = 7$, grey bars) and chaotic ($\omega = 0.5$, red line) phase spaces, compared to Poisonnian and circular orthogonal ensemble (COE) predictions for fully integrable and nonintegrable Hamiltonians, respectively. b) Average quasienergy level spacing parameter, $\avg{r}$, as a function of driving frequency, $\omega$, showing the transition from COE statistics to Poissonian statistics. c) Distributions of Shannon entropy, $\mathcal{S}$, of the effective Hamiltonian eigenstates in the Floquet basis. The Poissonian prediction is $P(\mathcal{S}) = \delta(\mathcal{S})$. d) Average Shannon entropy  $\avg{\mathcal{S}}$ as a function of modulation frequency. e) Distribution of FOTOCs sampled across phase space after $J_0t=30$, slightly after the Ehrenfest time. f) Average of FOTOCs at $J_0t=30$ as a function of modulation frequency. Parameters for all plots are $N = 1000$, $NU = -1$ and $J(t) = J_0 + 1.5\cos{(\omega t)}$.}
\end{figure}

The modulated dimer Hamiltonian~\eqref{eq:H} displays two symmetry classes, defined by the two possible eigenvalues, $\pm1$, of the parity operator, $\hat{P} = (-i)^N e^{-i\pi\hat{S}_x}$~\cite{Watanabe2012}. Floquet modes must be sorted into their respective symmetry classes before quasienergy level spacing statistics can be calculated. As shown in Fig.~\ref{fig:spectrum}a, the quasienergy level-spacing distribution for low driving frequency ($\omega = 0.5$) matches the circular orthogonal ensemble predictions, indicating nonintegrability~\cite{DAlessio2014}. For high driving frequency ($\omega = 7$), the quasienergy level spacing distribution matches the Poissonian distribution, indicating integrability~\cite{DAlessio2014}.

As the driving frequency is increased, we observe a transition from COE to Poissonian statistics. To pinpoint the critical driving frequency $\omega_c$ at which this occurs, we calculate the average level-spacing spacing parameter, defined as~\cite{DAlessio2014},
\begin{equation}
  r_n = \frac{\min{(\delta_n,\delta_{n+1})}}{\max{(\delta_n,\delta_{n+1})}} \in [0,1], \qquad
  \delta_n = \epsilon_{n+1} - \epsilon_n.
\end{equation}
The average level spacing is then given by $\avg{r} = \frac{1}{N} \sum_n r_n$. A high $\avg{r}$ indicates level repulsion and nonintegrability, while a low value indicates level clustering and integrability~\cite{DAlessio2014}. This is confirmed in  Fig.~\ref{fig:spectrum}b, where the $\avg{r}$ agrees with the COE prediction for low driving frequency, and rapidly falls away as as the driving frequency is increased into the regime where we expect the Magnus expansion to hold.

The signatures of this transition are also evident in the Shannon entropy of the system. The Shannon entropy indicates the average delocalisation of Floquet modes in the basis of effective Hamiltonian eigenstates. To produce comparable results to Figs.~\ref{fig:spectrum}a-b, we plot distributions of Shannon entropy for regular and chaotic phase space parameters in Fig.~\ref{fig:spectrum}c and the average Shannon entropy as function of modulation frequency in Fig.~\ref{fig:spectrum}d. In Fig.~\ref{fig:spectrum}d we show that the Shannon entropy rises as the modulation frequency is decreased. Unlike the level spacing parameter, however, the Shannon entropy never fully reaches reach the COE prediction over the frequency values plotted. This behaviour is due to the greater sensitivity of the Shannon entropy to finite-size effects as compared to the level-spacing parameter. Thus, we expect that far from the thermodynamic limit, it is only for very low modulation frequency that the Shannon entropy will reach the COE prediction~\cite{DAlessio2014, Santos2010_1, Santos2010_2}. %We expect that for sufficiently small modulation frequency, the Shannon entropy will saturate to the COE prediction.

For intermediate driving frequency, the phase space displays both integrable and chaotic regions and the level spacing distribution, a global indicator of chaos, lies between the Poissonian and COE predictions. To differentiate between chaotic and integrable phase space in this regime, we use a local indicator of chaos, the FOTOC.

To generate Figs.~\ref{fig:spectrum}e and f we calculate the long-time value of FOTOCs of coherent-state centred operators~\eqref{eq:local_op} distributed across the phase space in a 21 by 20 grid. For small modulation frequency, all phase space FOTOCs saturate to the infinite-temperature diagonal-ensemble prediction, as indicated by the small variance in the distribution of FOTOC values for $\omega = 0.5$ plotted in Fig.~\ref{fig:spectrum}e. The tendency of long-time FOTOC values to the infinite-temperature diagonal-ensemble prediction for small driving frequency is consistent with the Floquet eigenstate thermalisation hypothesis for interacting, isolated, periodically driven systems exhibiting thermalisation~\cite{Seetharam2018, Haldar2018}. For large modulation frequency $\omega = 7$ (grey solid distribution in Fig.~\ref{fig:spectrum}e) most FOTOCs retain a low value at $t = 100J_0$, but some grow larger than the diagonal ensemble prediction. The latter behaviour is due to the large, regular shearing of the phase-space distribution of some states over time. Nevertheless, the average FOTOC value as a function of modulation frequency shows a sharp transition between regular and chaotic regimes, as indicated in Fig.~\ref{fig:spectrum}f.

As the long-time FOTOC can attain large values for both chaotic and regular phase space, a large-valued FOTOC is not a sufficient indicator of quantum chaos or thermalisation. Quantum chaos can be identified from FOTOC dynamics by the presence of exponential growth up until the Ehrenfest time and saturation to the diagonal ensemble prediction, as indicated in Fig.~\ref{fig:FOTOC}. In the case of initial states with support on both regular and chaotic phase space regions, the short-time FOTOC dynamics may feature irregular growth, as shown in Fig.~\ref{fig:FOTOC2}. Therefore, choosing appropriate initial coherent state centre and width, which can be decreased by increasing particle number, $N$, is important when using FOTOCs to diagnose chaos and thermalisation.

\section{Conclusions}

In this work we explored the connections between classical chaos, the scrambling of quantum information, and the predictions of the eigenstate thermalisation hypothesis in Floquet systems.
In particular, we considered a driven Bose-Hubbard dimer model, which features a regular-to-chaotic transition in its semiclassical dynamics. We used Floquet analysis to study the various signatures of this transition in the quantum dynamics of the system, such as the exponential growth of the fidelity out-of-time-order correlator, from which a quantum Lyapunov exponent can be extracted.

Moreover, we were able to establish a link between the exponential growth of the FOTOC and thermalisation using several measures. First, we compared the saturation value of the FOTOC to the predictions of the periodic Gibbs ensemble, as required by the eigenstate thermalisation hypothesis. Next, we showed that this apparent thermalisation is also evident in the level spacing statistics. As the system transitions from integrable to chaotic, the level spacing statistics are no longer Poissonian but instead correspond to the circular orthogonal ensemble of random matrices. Finally, we showed that the divergence of the Floquet-Magnus perturbative expansion is another signature of the apparent thermalisation of Floquet system. We used the Shannon entropy to quantify the onset of this failure via the delocalisation of the Floquet  in the effective Hamiltonian eigenstate basis.

Our results build on, and are complementary, to previous studies in Refs.~\cite{Kidd2019, Lewis-Swan2019, DAlessio2014}. Our protocol is realisable experimentally and provides an alternative platform to explore ideas relevant to quantum thermodynamics and the dynamics of quantum information.

% \bibliographystyle{apsrev4-2}
% \bibliography{references.bib}


\end{document}